\documentclass{article}

\usepackage{arxiv}

\usepackage[utf8]{inputenc} 
\usepackage[T1]{fontenc}    
\usepackage{hyperref}       
\usepackage{url}            
\usepackage{booktabs}       
\usepackage{amsfonts}       
\usepackage{nicefrac}       
\usepackage{microtype}      
\usepackage{graphicx}
\usepackage{natbib}
\usepackage{doi}

\title{Ruled by the Representation Space: On the University’s Embrace of Large Language Models}

\date{May 6, 2025}

\author{ \href{https://orcid.org/0000-0002-7938-2608}{\includegraphics[scale=0.06]{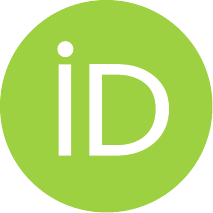}\hspace{1mm}Katia Schwerzmann}\thanks{This contribution will appear in German in (Breil and Sprenger 2025). Thank you to Deanna Cachoian-Schanz, Kerry McAuliffe, Florian Sprenger, and Patrizia Breil for their thoughtful reading.} \\
	SFB Virtuelle Lebenswelten\\
    Ruhr-Universität Bochum, Germany\\
	\texttt{katia.schwerzmann@rub.de} \\
}

\renewcommand{\headeright}{Preprint}
\renewcommand{\undertitle}{Preprint}
\renewcommand{\shorttitle}{Ruled by the Representation Space}

\hypersetup{
pdftitle={Ruled by the Representation Space},
pdfsubject={cs.CY},
pdfauthor={Katia Schwerzmann},
pdfkeywords={Representation Space, LLM, Generative AI, Virtual, University},
}

\begin{document}
\maketitle

\begin{abstract}
	{This paper explores the implications of universities’ rapid adoption of large language models (LLMs) for studying, teaching, and research by analyzing the logics underpinning their representation space. It argues that by uncritically adopting LLMs, the university surrenders its autonomy to a field of heteronomy, that of generative AI, whose norms are not democratically shaped. Unlike earlier forms of rule-based AI, which sought to exclude human judgment and interpretation, generative AI’s new normative rationality is explicitly based on the automation of moral judgment, valuation, and interpretation. By integrating LLMs into pedagogical and research contexts before establishing a critical framework for their use, the University subjects itself to being governed by contingent, ever-evolving, and domain-non-specific norms that structure the model’s virtual representation space and thus everything it generates.}
\end{abstract}


\section{The University’s Autonomy Must Be Defended}
{This text addresses the sweeping acceptance by German Universities of the use of large language models (LLMs) for the purpose of teaching and research: “A complete ban on generative AI does not seem feasible in general and does not make sense at all \textit{if AI applications will become part of a professional conduct in science and other fields} [emphasis mine]” (Ruhr Universität Bochum 2025).\footnote{At the Ruhr-Universität Bochum, extensive training is offered to professors and lecturers regarding the implementation and use of LLMs and generative AI in both teaching and research (Ruhr Universität Bochum 2025). The RUB even offers its own “privacy friendly” version of GPT (ebd.). The evaluation of students’ work with the help of AI is authorized as long as the result of the evaluation is “critically reviewed,” adding that the “grade can only be determined by the person responsible for the evaluation” (ebd.) In the US, OpenAI offers expensive versions of its tools for research purposes. The University of California, among others, has partnered with OpenAI to roll out an “education-specific version” of GPT (Kant 2025).} Acknowledging the ways in which the virtual is “world-making” and rejoining Patrizia Breil and Florian Sprenger’s invitation to contribute to an idea of the University as a place for thinking “beyond what is given” (2025, Introduction), I ask under what conditions the University can preserve its openness to that which is not simply “given”; such are possible futures, new fields of knowledge, new forms of study, new relationships with its members and with society. Here, the “given” is not only that which is considered as being presently the case but also that which is believed to be the case in the future. What is considered a “given” lays a normative claim to the future by which a possibility is conjured up as the reality (\textit{“if . . . will become”}) in the name of which the University authorizes present practices as irremissible: “A complete ban on generative AI . . . does not make sense at all if AI applications will become part . . . ” (Ruhr Universität Bochum 2025). In its somewhat awkward phrasing, the University reveals a deeper truth: the “given” is never merely a natural or actual fact. Rather, it is shaped by \textit{what should be} precisely because \textit{it will be}—a form of determinism that goes hand in hand with a relinquishment of autonomy and responsibility.

First hypothesis: the University can preserve its openness “beyond what is given” under the condition that it remains able to produce, at least in part, its own rules. Autonomy—the ability to set one’s own rules—is never absolute. It is relational, negotiated in response to the heteronomy of other fields. In the case of the University, these other fields are the political and the economic. I argue that renouncing autonomy, even as a regulative ideal, amounts to surrendering to the “given” and thus, renounces its ability to produce rules,  forsaking the virtual construed as that which is “beyond the given.” In other words, to reject autonomy—even as an aspirational principle—is to accept the world as merely what is, while abandoning the ability to shape what could be.

What happens when the University throws itself into a field of heteronomy whose “rules” are ever shifting, implicit, and escape being challenged or actively shaped? This field in question is none other than the field of machine-learning-based generative AI, and more precisely, large language models (LLMs). LLMs are machine learning models trained on big datasets that generate output by predicting the sequence of tokens that have the statistically highest likelihood of following the words in the user’s input. After the model has been trained to capture statistical patterns present in the training dataset, it is able to encode the user’s input while also considering the relative importance of each token and its relationships to surrounding ones.\footnote{This process is commonly referred to as “attention mechanism” (Vaswani et al. 2017) and is one of the reasons for the success of LLMs.} 

The generative character of this type of AI relies on the construction of a virtual “representation space” that is not only a representation of what of the world matters (Amoore et al. 2024); it is also world-making as a result of its specific normativity, which appears legitimate as an effect of machine learning’s “artificial naturalism” (Campolo and Schwerzmann 2023, 2–3). This artificial naturalism consists in the belief that world knowledge is “reflected” in the large dataset and accurately captured by the model during training. Consequently, the models’ output takes on an effect of “givenness” that makes it difficult to challenge. 

What LLMs’ representation space forecloses is what I propose to call generative AI’s \textit{normative rationality}. Second hypothesis: this normative rationality is the new type of algorithmic rationality that marks a stark departure from earlier forms of rule-based artificial intelligence. In their book \textit{How Reason Almost Lost Its Mind,} historian Erickson and colleagues frame these earlier forms of AI within a “Cold War Rationality”—a rationality devoid of human judgment and interpretation and thus considered particularly apt to minimize uncertainty (Erickson et al. 2013; Daston 2022). By contrast, current machine-learning-based AI explicitly relies on the automation of moral judgement, valuation, and interpretation. While machine learning models reflect or express statistical regularities present in the dataset, this reflection is normative for several reasons: first, the training data embodies valuations shaped by specific configurations of power within the society from which it originates; second, large datasets undergo both \textit{curation} and \textit{feature engineering} to optimize models’ ability to learn from them—a process of “exemplification” (Campolo and Schwerzmann 2023); third, generative models undergo a second training phase called fine-tuning, the purpose of which is to regulate models’ “undesirable behaviors” (Ouyang et al. 2022). Fine-tuning consists in “aligning” models with a set of values that reflect what big tech companies and national actors consider “preferred” behaviors (Schwerzmann and Campolo 2025). It involves introducing a more explicit type of normativity into the model  by retraining it on a small number of human-written and -evaluated examples embodying desirable values—a costly process called reinforcement learning. 

Following Foucault, “norms” are both a means of directing conduct—“the conduct of conduct” (Foucault 2008, 186)—and representations of what society could or should be (Foucault 2018, 214). Norms not only stabilize social orders but also play a crucial role in shaping new orders and forms of sociality. Norms embody values, which, following Nietzsche, are qualitative appraisals of the world as something that matters to us (Nietzsche 1974, 241–42). AI’s normative rationality presents an image of what the makers of models desire for society (Schwerzmann and Campolo 2025).}

\section{Note on the Virtual}
{How is the virtual to be understood here? The virtual is as real in its effects as the actual. In his discussion of Bergson, Proust, and Deleuze, Rob Shields calls the virtual a “real idealization” and describes it as all that is “almost-so” (Shields 2005, 28). Like the actual, the virtual affects the world in a physical way. To clarify the relation between the virtual and the actual, I understand the virtual as the latent, that is, that which is at the periphery of the actual and from which the actual sometimes crystallizes. I propose that the relation of the virtual to the actual can be \textit{quantitative} or \textit{qualitative} in nature. Quantitatively, the virtual differs from the actual in terms of likelihood and probabilities for something to happen compared to what has actually happened. Construed in this way, the virtual can be a space of modeling of, experimentation on, and extrapolation from the real. The quantitative virtual is tethered to the “given”—the indexical character of the data (Kitchin and McArdle 2016)—as that which constitutes its norm. When qualitative, the virtual differs in nature from the actual and is irreducible to it. (Science-)Fiction, speculation, games—all opening a plurality of worlds untethered to the “given”—are of the qualitative kind of virtual.\footnote{This distinction between a quantitative and a qualitative virtual corresponds to the distinction between (science-)fiction and extrapolation on which Ursula K. Le Guin insists in her introduction to (Le Guin 2019).} The qualitative virtual may incline us to challenge the “given” by helping us to imagine what could be.

As stated at the outset, the Virtual University is one that keeps open the possibility of a plurality of worlds, refusing to uncritically yield to the heteronomy of neighboring fields or to the “given.” Yet the University has very quickly and uncritically embraced LLMs and with them, their “representation space.”}

\section{The Virtual of Generative AI}
{LLMs, like other generative models, operate based on a high-dimensional space that consists of representations that the model captures from the training data.\footnote{In LLMs this representation space is called “vector space,” while in image generation models it is called “latent space.” For a critical analysis of GANs’ latent space, see (Offert 2021).} This vector space enables and constrains what the model generates. Following the typology proposed above, the virtuality of this representation space is quantitative: the model maps and compresses in this lower dimensional space the probability distribution of the “input space” made of the actual data (Bengio, Courville, and Vincent 2013). 

When the model generates content, it actualizes an area of the representation space—an operation called “interpolation” (Arvanitidis, Hansen, and Hauberg 2021; Chollet 2021). The fact that generative models interpolate rather than extrapolate is significant: so far, generative models do not generate output beyond the probability distribution of the actual data (which would be extrapolation) but instead “sample” the representation space between actual data points (which is \textit{inter}polation). Therefore, models’ ability to process not-yet-seen data happens within the bounds of the data they have already “seen.” The “dense” regions of the representation space correspond to areas where the model has learned more representations, resulting in higher output certainty compared to sparser regions (Arvanitidis, Hansen, and Hauberg 2021, 4). When prompted outside of these dense zones, the model tends to “hallucinate” and produce inaccurate results (Ji et al. 2023). “Prompt engineering”’s purpose is to turn the model’s “attention” to the more densely populated areas of the representation space, which allows for a “better” encoding of the prompt.

Following the typology I propose, the virtuality of the representation space is of a quantitative nature, since everything the model generates remains tethered to actual data.}

\section{The Normativity of The Representation Space}
{By embracing LLMs before developing any critical framework for their use in pedagogical and research contexts, the University allows itself to be governed by the contingent, ever-evolving, implicit, and domain-non-specific norms that shape the model’s virtual representation space—norms reflecting the ideology of the now Trump-aligned Silicon Valley tech sector or Chinese state ideology, like in the case of DeepSeek. With their claim to being “neutral and balanced,” LLMs not only systematically obfuscate the ethico-technical work needed to produce this claim of normative neutrality; they also neutralize the question of positionality and of partial and situated knowledge—a major epistemological contribution of feminist and postcolonial studies—by claiming a gaze from everywhere that aspires to capture everything and represent everything equally (Schwerzmann 2024).

Additionally, by accepting generative models’ normative rationality, the University reframes learning and research as essentially consisting in the evaluation of models’ outputs. Instead of teaching students to become the active producers of meaning, analysis, and arguments, students learn to become the evaluators, modulators, and improvers of synthetic output. However, these evaluations are based on and therefore determined by the domain-non-specific normative rationality of generative models—a rationality that is left unchallenged by this optimization process. Additionally, by serving as the evaluative supplement to models’ output, researchers forfeit the idiosyncratic and creative dimension of knowledge production, the ability to invent new approaches and new forms of knowledge. 

At the beginning, I defined the “given” as simultaneously a current and an expected state of the world determined by technology. This definition must be expanded: the “given” is both an \textit{effect} and \textit{affect} of givenness generated by current AI’s artificial naturalism and its specific normative rationality. If the University is to resist surrendering entirely to the “given” and remain a space of experimentation, of not only quantitative but also qualitative virtuality, it must defend itself—through a critical analysis that accounts for the logics of technology and its material-discursive effects—against its subordination to the heteronomy of AI’s virtual representation space; a space ruled by unchallenged, naturalized norm-giving operations.}

\bibliographystyle{unsrtnat}
\nocite{*}
\bibliography{references}
{Amoore, Louise, Alexander Campolo, Benjamin Jacobsen, and Ludovico Rella. 2024. “A World Model: On the Political Logics of Generative AI.” Political Geography 113:1–9.

Arvanitidis, Georgios, Lars Kai Hansen, and Søren Hauberg. 2021. “Latent Space Oddity: On the Curvature of Deep Generative Models.” arXiv. https://doi.org/10.48550/arXiv.1710.11379.

Bengio, Yoshua, Aaron Courville, and Pascal Vincent. 2013. “Representation Learning: A Review and New Perspectives.” IEEE Transactions on Pattern Analysis and Machine Intelligence 35 (8): 1798–1828. https://doi.org/10.1109/tpami.2013.50.

Breil, Patrizia, and Florian Sprenger, eds. 2025. Virtuelle Universität. Virtuelle Lebenswelten. Bielefeld: Transcript.

Campolo, Alexander, and Katia Schwerzmann. 2023. “From Rules to Examples. Machine Learning’s Type of Authority.” Big Data and Society 10 (2): 1–13. https://doi.org/10.1177/20539517231188725.

Chollet, François. 2021. Deep Learning with Python. Shelter Island, NY: Manning Publications Co.

Daston, Lorraine. 2022. Rules: A Short History of What We Live By. Princeton: Princeton University Press.

Erickson, Paul, Judy L. Klein, Lorraine Daston, Rebecca Lemov, Thomas Sturm, and Michael D. Gordin. 2013. How Reason Almost Lost Its Mind: The Strange Career of Cold War Rationality. Chicago/London: The University of Chicago Press.

Foucault, Michel. 2008. The Birth of Biopolitics: Lectures at the Collège de France, 1978–1979. Translated by Graham Burchell. New York: Picador.

———. 2018. The Punitive Society: Lectures at the College de France, 1972-1973. Edited by Bernard E. Harcourt. Translated by Graham Burchell. New York: Picador.

Ji, Ziwei, Nayeon Lee, Rita Frieske, Tiezheng Yu, Dan Su, Yan Xu, Etsuko Ishii, Ye Jin Bang, Andrea Madotto, and Pascale Fung. 2023. “Survey of Hallucination in Natural Language Generation.” ACM Computing Surveys 55 (12): 1–38. https://doi.org/10.1145/3571730.

Kant, Rishi. 2025. “OpenAI Targets Higher Education in the U.S. with ChatGPT Rollout at California State University.” Reuters, February 4, 2025, sec. Technology. https://www.reuters.com/technology/openai-targets-higher-education-us-with-chatgpt-rollout-california-state-2025-02-04/.

Kitchin, Rob, and Gavin McArdle. 2016. “What Makes Big Data, Big Data? Exploring the Ontological Characteristics of 26 Datasets.” Big Data and Society 3 (1): 1–10. https://doi.org/10.1177/2053951716631130.

Le Guin, Ursula K. 2019. The Left Hand of Darkness (1969). New York, NY: Ace Books.

Nietzsche, Friedrich. 1974. The Gay Science: With a Prelude in German Rhymes and an Appendix of Songs. Translated by Walter Kaufmann. New York: Random House. https://doi.org/10.1017/CBO9780511812088.

Offert, Fabian. 2021. “Latent Deep Space: Generative Adversarial Networks (GANs) in the Sciences.” Media+Environment, 1–20. https://doi.org/10.1525/001c.29905.

Ouyang, Long, Jeff Wu, Xu Jiang, Diogo Almeida, Carroll L Wainwright, Pamela Mishkin, Chong Zhang, et al. 2022. “Training Language Models to Follow Instructions with Human Feedback.” arXiv, 1–68. https://doi.org/arXiv:2203.02155.

Ruhr Universität Bochum. 2025. “Künstliche Intelligenz in Studium und Lehre.” Zentrum für Wissenschaftsdidaktik. 2025. https://zfw.rub.de/lehrende/lehre-gestalten/kuenstliche-intelligenz-in-studium-und-lehre/.

Schwerzmann, Katia. 2024. “From Enclosure to Foreclosure and Beyond: Opening AI’s Totalizing Logic.” Philpapers (Preprint). https://philpapers.org/rec/SCHFET-6.

Schwerzmann, Katia, and Alexander Campolo. 2025. “‘Desired Behaviors’: Alignment and the Emergence of a Machine Learning Ethics.” AI and Society, 1–14. https://doi.org/10.1007/s00146-025-02272-3.

Shields, Rob. 2005. The Virtual. Hoboken: Taylor and Francis.

Vaswani, Ashish, Noam Shazeer, Niki Parmar, Jakob Uszkoreit, Llion Jones, Aidan N. Gomez, Lukasz Kaiser, and Illia Polosukhin. 2017. “Attention Is All You Need.” arXiv, December. http://arxiv.org/abs/1706.03762.}
\end{document}